\documentclass[sigconf]{acmart}

\usepackage{multirow}
\usepackage{pifont}% http://ctan.org/pkg/pifont
\usepackage{wrapfig}
\usepackage{subcaption}
\newcommand{\cmark}{\ding{51}}%
\newcommand{\xmark}{\ding{55}}%
\AtBeginDocument{%
  \providecommand\BibTeX{{%
    \normalfont B\kern-0.5em{\scshape i\kern-0.25em b}\kern-0.8em\TeX}}}

\setcopyright{acmcopyright}
\copyrightyear{2022}
\acmYear{2022}
\acmDOI{10.1145/1122445.1122456}

\acmConference[HILDA '22]{}{2022}{XXXX, XXX}
\acmPrice{15.00}
\acmISBN{978-1-4503-XXXX-X/18/06}

\begin{document}

\title{HumanAL: Calibrating Human Matching Beyond a Single Task}

\author{Roee Shraga}
\email{r.shraga@northeastern.edu}
%\email{webmaster@marysville-ohio.com}
\affiliation{%
  \institution{Northeastern University}
%  \streetaddress{P.O. Box 1212}
  \city{Boston}
  \state{MA}
  \country{USA}
}
\begin{abstract}

This work offers a novel view on the use of human input as labels, acknowledging that humans may err. We build a behavioral profile for human annotators which is used as a feature representation of the provided input. We show that by utilizing black-box machine learning, we can take into account human behavior and calibrate their input to improve the labeling quality. To support our claims and provide a proof-of-concept, we experiment with three different matching tasks, namely, schema matching, entity matching and text matching. Our empirical evaluation suggests that the method can improve the quality of gathered labels in multiple settings including cross-domain (across different matching tasks).

\end{abstract}
\maketitle

\section{Introduction}\label{sec:intro}
The basic setting of machine learning, supervised learning, provides an algorithm with a set of human-labeled data instances from which it can ``learn'' a model. Nevertheless, for many tasks, relying on human-labeled examples to support learning may be problematic due to several reasons that include, among others, the existence of label noise~\cite{frenay2013classification,northcutt2021pervasive}, and the costly acquisition of high-quality labels~\cite{sheng2008get}. Consequently, human-in-the-loop solutions~\cite{li2017human,NoyMMA13,Crowdmap,fan2014hybrid,McCann2008,Hung2014,pinkel2013incmap,zhang2019invest,lertvittayakumjorn2020find}, such as crowdsourcing, were developed to enable cheaper acquisition of labels from a large number of annotators in a short time. However, the increasing availability of human-labeled data may also yield variation in the reliability and proficiency of these human annotators~\cite{dragisic2016user,zhang2018reducing,Ross2010,HILDA18,ackerman2019cognitive,shraga2020learning}, which, in turn, impair the quality of the generated labels. This variation may arise due to different expertise levels~\cite{dragisic2016user,zhang2018reducing,HILDA18}, mood changes~\cite{DBLP:conf/vldb/Shraga20}, various cognitive biases~\cite{ackerman2019cognitive,shraga2020learning} and more. 

In this work we offer a novel view on the use of human input as labels, acknowledging that humans may err. Building on top of meta-cognition research, we adopt cognitive biases that may affect the ability of humans to provide accurate answers. For example, previous research, in the context of matching, shows that decision times are associated with the decision confidence and correctness~\cite{ackerman2019cognitive}. In what follows, we build a behavioral profile for human annotators' decisions which is used as a feature representation of the provided input. We show that utilizing black-box machine learning, we can take into account human behavior and calibrate their input to improve the labeling quality. To support our claims and provide a proof-of-concept, we use matching tasks of three different flavors. Specifically, we perform an empirical study with the structurally rich database task of schema matching (SM)~\cite{RAHM2001,BELLAHSENE2011}, the structured task of entity matching (EM)~\cite{elmagarmid2006duplicate,getoor2012entity,christen2012data} and the textually rich NLP task of labeling semantically equivalent sentences (text matching, TM)~\cite{pang2016text,guo2019matchzoo}. SM is characterized as a very challenging task~\cite{shraga2020learning} and was shown in literature to require expert skills~\cite{shraga2020learning,zhang2018reducing,dragisic2016user}. EM usually depends on context and may be confusing for humans (e.g., what it means to be a match)~\cite{doan2017human,doan2018human}. For NLP tasks, human performance (aka ``Human Baselines'') is usually used as the verge of success for machine-based methods~\cite{nangia2019human,wang2019superglue}. We demonstrate that using our methodology we can improve (raw) human performance by acknowledging their imperfection to provide valuable labels.     

\begin{figure}[h]
	\centering
	\includegraphics[width=.5\textwidth]{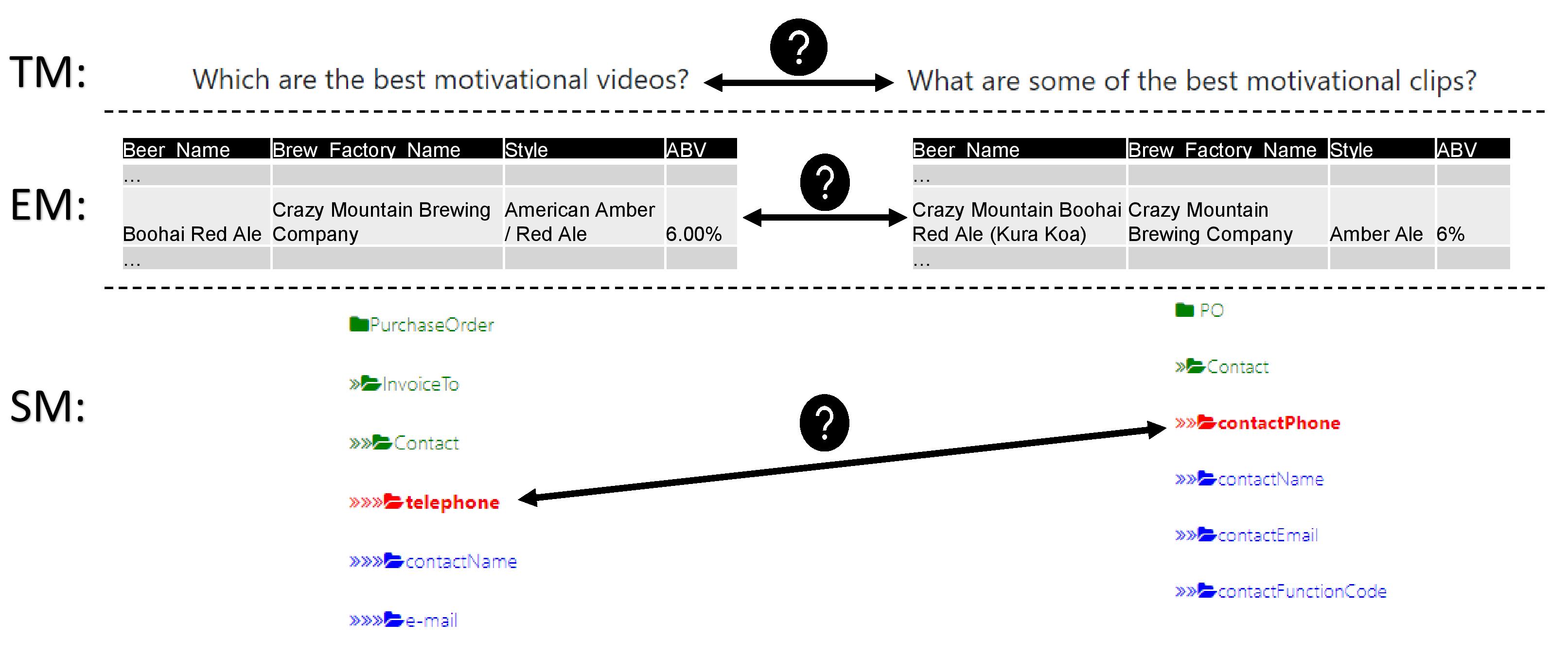}
	\caption{Example of each task that we consider in this paper, namely Text Matching (TM, top), Entity Matching (EM, middle), and Schema Matching (SM, bottom).}
	\label{fig:example}
\end{figure}

Matching can be seen as a classification problem of pairs of elements (sentences for TM, entities for EM, and attributes for SM). In the context of this paper, we consider that these tasks are provided for human matchers either as labelers or additional problem solvers in the loop. Figure~\ref{fig:example} provides an example sample for each task. At the top, we illustrate an example of TM, specifically taken from the quora question pairs matching dataset~\cite{qqp}, a well-known task from the GLUE benchmark~\cite{wang2018glue}\footnote{GLUE (\url{https://gluebenchmark.com/}) is a standard collection of benchmarks for natural language understanding, widely used in the NLP community.}. TM tasks are usually provided without context, i.e., each sentence pair is evaluated independently. Below the TM example, an EM example is provided taken from Magellan's~\cite{konda2016magellan} Beer dataset~\cite{beer}. In EM, the context of an entity pair usually includes its schema (in this case, \texttt{Beer\_Name, Brew\_Factory\_Name, Style, ABV}) and, sometimes, additional entities in the table. At the bottom, a SM task example is provided, for which, as illustrated, the autonomous attribute alignment task (\texttt{telephone} $\leftrightarrow$ \texttt{contactPhone}) is provided within the context of their respective schemata.

\subsection{Contributions}

This paper presents an ongoing study of the way we calibrate human input to generate HumanAL (\textbf{Human} \textbf{A}ugmented \textbf{L}abels). The main take-away is, that by training a calibration model on a labeled matching benchmark, we can improve the quality of (unknown) labelers on new (unseen) benchmarks from multiple domains. Our main contributions include:
\begin{enumerate}
	\item A novel behavioral profile of human matchers decisions in a crowdsourcing setting.
	\item A machine-learning based calibration model to improve the quality of labels for matching tasks. 
	\item An empirical study using over 300 participants providing 12,000 matching decisions ranging over three different matching tasks using well known benchmarks showing the effectiveness of our approach.%\footnote{The gathered data and code will become publically available upon acceptance}  
\end{enumerate}

In Section~\ref{sec:con} we provide a discussion regarding the current state of research and elaborate on future directions.

\section{Related Work}\label{sec:related}

Matching works over the years assume superiority of humans over algorithmic matchers. Traditional crowdsourcing and pay-as-you-go frameworks use humans in-the-loop to validate algorithmic matching solutions~\cite{wang2012crowder,gokhale2014corleone,Crowdmap,zhang2018reducing,dragisic2016user,li2019user,zhang2019invest,lertvittayakumjorn2020find}. Others aim to improve the quality of human labelers~\cite{shraga2021powarematch,yang2018cost}. Alternative research directions aim to surpass the need for humans in-the-loop by introducing unsupervised learning solutions~\cite{kolyvakis2018deepalignment,zhang2020unsupervised,wu2020zeroer} and transfer learning methods~\cite{jin2021deep,shraga12020} or limiting the amount of labels for training using active learning mechanisms~\cite{meduri2020comprehensive,kasai2019low,jain2021deep} and few-shot learning~\cite{kasai2019low,ye2019multi}. In this work we analyze and aim to improve the quality of labels provided in a crowdsourcing setting. Also, we analyze more than one matching task under the same methodology.

More general literature on assessing human expertise and quality suggest to identify low quality crowd workers~\cite{callison2009fast,ipeirotis2010quality} %and computer user skill~\cite{ghazarian2010automatic} 
mainly relying on gold questions to infer quality in practice~\cite{daniel2018quality}. Similar to our approach, Rzeszotarski and Kittur~\cite{rzeszotarski2011instrumenting} and follow up works~\cite{wu2016novices,goyal2018your} feature engineer human behavior to assess quality. Different from these approaches, we use the extracted features, not only to understand and assess, but also to improve the quality of labels for matching tasks.

Recently, Ackerman et al. detected human cognitive biases affecting matching quality~\cite{ackerman2019cognitive}. These biases were used in the scope of open-ended crowdsourcing~\cite{parameswaran2016optimizing} for the task of improving human schema matching~\cite{shraga2020learning,shraga2021powarematch}. The current work focuses on boolean crowdsourcing aiming to address multiple human matching tasks.
\section{Learning to Calibrate Human Input}\label{sec:modelMain}

The setup for this paper is \emph{boolean crowdsourcing}, i.e., a user is required to solve a sequence of binary matching questions. While this is a typical setup for multiple human-in-the-loop matching~\cite{li2017human}, we now focus on human labelers. %for labeling. 
We model human %begin by modeling this 
input and provide a feature set to be used to calibrate these binary decisions and generate new human augmented labels (HumanAL).

\subsection{Modeling Human Input}\label{sec:model}

We consider a machine learning-like model for labeling defined as follows. Let $\mathcal{X}$ be the space of samples. For the scope of this paper, a sample $x\in \mathcal{X}$ represents a pair of elements, \emph{e.g.,} pair of sentences. We further denote by $X = \{x_1, x_2, \dots\}$ a sequence of samples.\footnote{we use a \emph{sequence} of samples rather than \emph{set} of sample since in this setting the samples are presented to a user in a given order.} We assume that there exists some ``true'' label (sometimes referred to as ground truth) for each sample in the space $\mathcal{X}$, in a space $\mathcal{Y}$. Our objective is classify pairs of elements (samples) as match (1) or non-match (0) and, thus, consider $\mathcal{Y} = \{0,1\}$. The sequence $Y = \{y_1, y_2, \dots\}$ denotes $X$'s respective labels. Finally, a labeled dataset $D = \{(x_1, y_1), (x_2, y_2), \dots\}$ is a sequence of pairs $(x_i, y_i)$ such that $x_i$ is a sample and $y_i$ is its label (denoting whether the pair matches, i.e., $y_i = 1$ or not, i.e., $y_i = 0$).

In a traditional setting, the construction of a labeled dataset $D$ involves \emph{human labelers} and it is usually assumed that given an instance $x$, the label provided by a labeler is \emph{correct}. Instead, we view human labeling as \emph{predictions} and use the notation $\hat{y}_i$ for human input regarding a sample $x_i$. In a perfect world, where humans do not err, we can simply set $y_i = \hat{y}_i$ assuming that the provided label is always correct. Alternatively, we can also include multiple labelers for each sample and use a majority voting mechanism. 

When dealing with human labelers in a real (not perfect) world, our objective is to utilize $\hat{y}_i$ and the properties of human decision making. Each decision regarding a sample involves additional essential components, for example, the \emph{decision time}, which was shown to be predictive for both confidence and accuracy in matching~\cite{ackerman2019cognitive}. In addition, instead of solely requesting a binary decision from a labeler, we also query their \emph{confidence level} regarding the sample~\cite{dragisic2016user}. In what follows, instead of viewing the human provided label $\hat{y}_i$ as a binary indicator, we view it as a triplet $\hat{y}_i = \langle l_i, c_i, t_i\rangle$, where $l_i\in \mathcal{Y}$ is the provided label for $x_i$, $c_i\in {\rm I\!R}$ is the associated confidence, and $t_i\in{\rm I\!R}$ is the decision time. Next, we show how this representation is used for feature extraction.  

%\subsection{From Human Behavior to Feature Representation}\label{sec:features}

\subsection{From Human Behavior to Feature Extraction to Calibration}\label{sec:ML}

\begin{figure}[t]
	\centering
	\includegraphics[width=.45\textwidth]{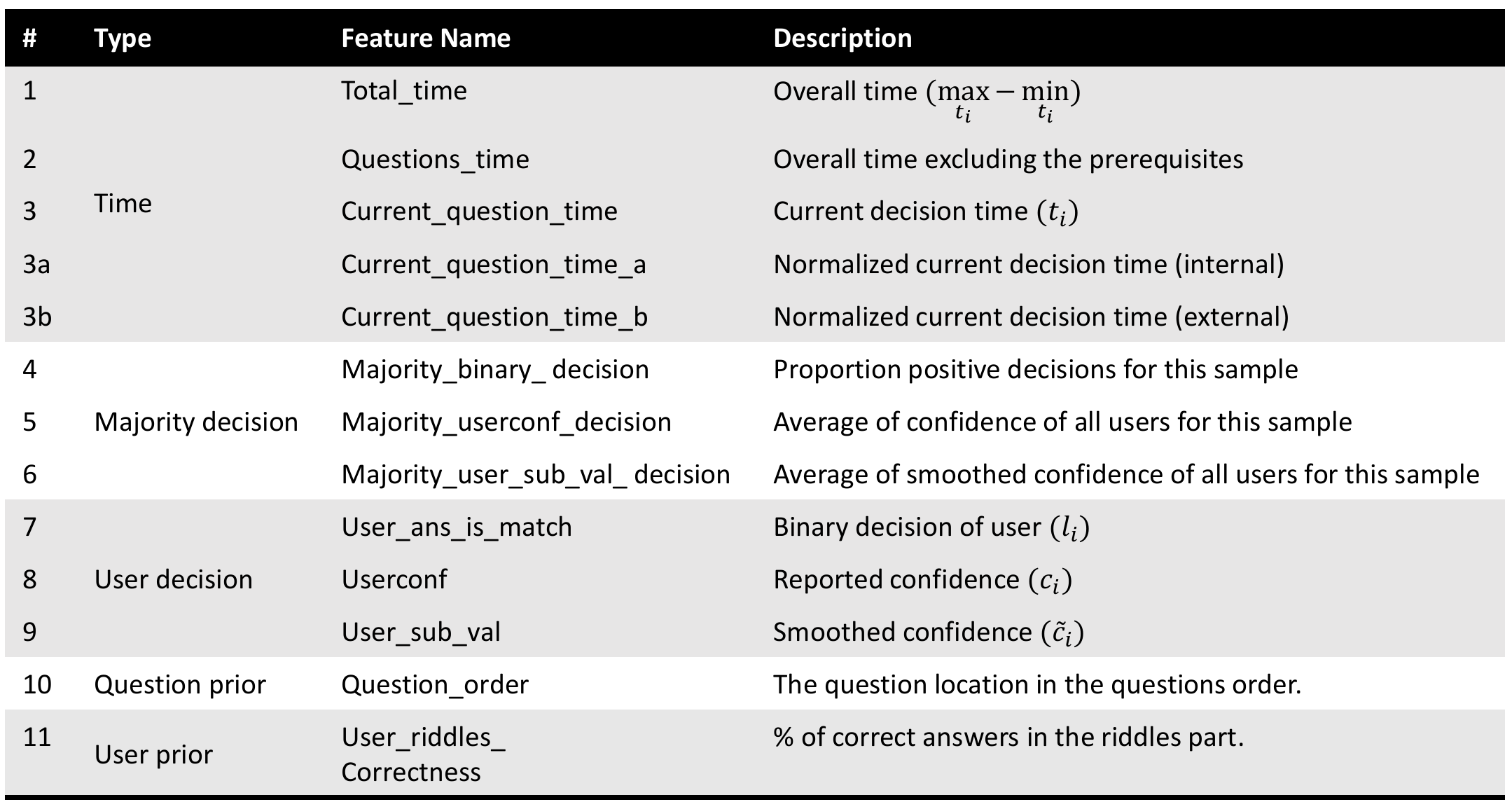}
	\caption{A behavioral profile of a human matcher decision}
	\label{fig:features}
\end{figure}

In this work, we use the aforementioned triplets to extract five classes of features which associate to human decision making. The full list of generated features is provided in Figure~\ref{fig:features}. The clear extraction that consists of the decision time $t_i$, binary decision $l_i$ and reported confidence $c_i$ can be seen as \emph{local} (extracted only from the current decision of the user) and \emph{internal} (determined only based on the user behavior). Another local internal feature we consider is the smoothed confidence, defined as follows.

\begin{equation}\label{eq:sconf}
	\tilde{c}_i = \begin{cases}
				  2\cdot(c_i-0.5) & c_i \geq 0.5 \\
				  2\cdot(0.5-c_i) & c_i < 0.5 
				  \end{cases}
\end{equation} 

\noindent The idea of the smoothed confidence is to encode the decisive nature into the confidence by dividing the scale into match/no-match ([0,0.5)/[0.5-1.0]). For example, when comparing a reported confidence of $0.7$ (associated with a match decision) to the one of $0.1$ (associated with a no-match decision), the latter, while lower in actual value, represents a higher confidence in the decision. In this case the respective smoothed confidences will be $0.4$ ($(0.7-0.5)\cdot 2$) and $0.8$ ($(0.5-0.1)\cdot 2$), respectively, suggesting that the latter decision (saying no-match) is made with higher confidence than that of the former (saying match). Finally, aiming to address the fatigue in sequential making, we also include the position of the currect decision in the pool of decisions provided to the user. For example, the confidence provided to the first seen sample might be different than the one reported for the tenth decision, even if they reflect the same internal confidence.

We also consider \emph{external} local features which use information regarding other users facing the same matching decision. Specifically, we normalize the decision time with respect to the decision times of other users over the same decision. In addition, all majority decision features are external and local with respect to the user decision features. For example, the average reported confidence of other users for the same matching decision. 

The overall time and (internal) normalized decision time with respect to other decisions made by the user are internal \emph{global} features. This set also includes a prior intelligence estimation using psychometric riddles~\cite{anastasi1997psychological,shraga2020incognitomatch}.

%Another classification of the feature set can be to global (extracted from a set of decisions made by the user) and local (extracted from a single decision made by the user) features and internal (determined over based on the user behavior) and external (determined over the user behavior with respect to other users) features. For example, ...

Using the generated features, we use a black-box machine learning strategy. Specifically, using the feature extraction, each user decision is now encoded into a feature space over which we wish to determine a new human augmented label. To do so, we position the problem as a classification problem, aiming to classify the new representation to a match (1)/no-match (0) decision. During training, we assume that there exists some ground truth label $y_i$ over which we train a classifier to map the suggested features into a human augmented label $\hat{y}^{*}_i$. 
\section{Empirical Evaluation}

We now describe our empirical evaluation which provides the following main take-aways: 
\begin{enumerate}
	\item The suggested behavioral profile (Section~\ref{sec:modelMain}), aided by a pre-trained black-box machine learning, improves label quality of text matching, schema matching and entity matching.
	%	\item The most significant improvement is when the training set of human matchers performed a the same matching task (same domain) over a different set of samples (V3).
	\item Majority decision and the user decision (Figure~\ref{fig:features}) are the most significant features.
	\item Different from the observations in the setting of open-ended crowdsourced matching~\cite{ackerman2019cognitive,shraga2020learning,shraga2021powarematch}, decision times are less significant in boolean crowdsourcing.  
\end{enumerate}

\subsection{Experimental Settings}\label{sec:setup}
We now describe the interface and specific matching tasks we use.

\subsubsection{Interface}\label{sec:incognito}
\begin{figure}[h]
	\centering
	\includegraphics[width=.5\textwidth]{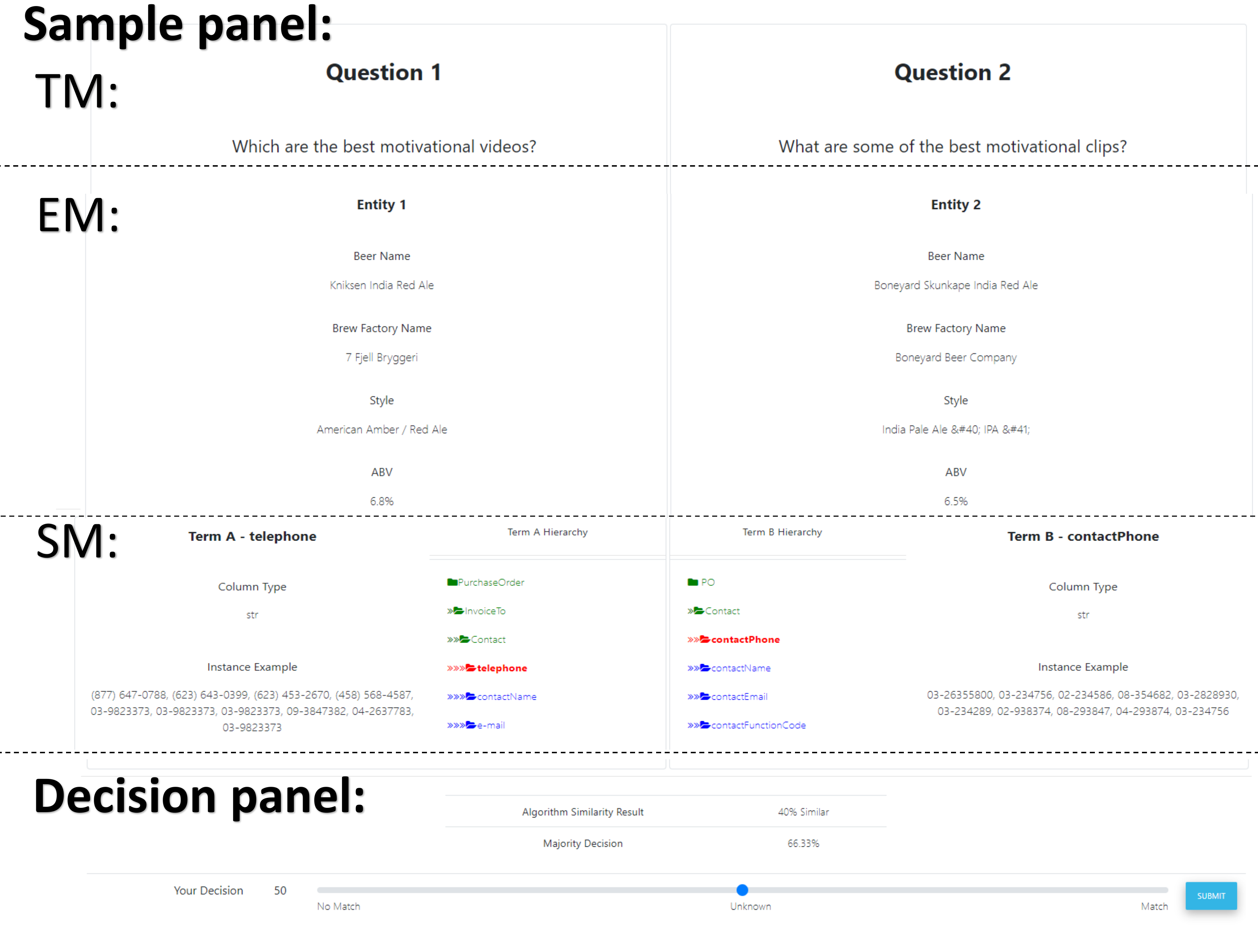}
	\caption{IncognitoMatch Interface for each Task. The sample panel, which is different for each task, is illustrated at the top. The decision panel, same for all tasks, is given at the bottom.}
	\label{fig:inteface}
\end{figure}

We adopt the interface of IncognitoMatch~\cite{shraga2020incognitomatch}, which was originally designed for schema matching problems. IncognitoMatch is a crowdsourcing platform that also gathers cognitive information such as decision times. Confidence is reported on the scale from 0 to 100 (which, in our setting, is normalized to scale between 0 and 1) such that 0 means fully confident no match, 50 means unknown and 100 means fully confident match (see Eq.~\ref{eq:sconf} for a smoothed variation of confidence). Please refer to the demonstration paper~\cite{shraga2020incognitomatch} for additional details. As each task differs in the context it provides, we created a different interface variation for each task. Figure~\ref{fig:inteface} provides screenshots from IncognitoMatch. The sample (element pair) panel for each task is given on top. The decision panel, illustrated at the bottom, is the same for all tasks.

\subsubsection{Matching Tasks Datasets}\label{sec:matching}
For the current study we use three datasets each representing a different matching task. The tasks are publicly accessible.

\noindent\textbf{Schema Matching:}\footnote{\url{https://agp.iem.technion.ac.il/incognitomatch/SchemaMatching/}} For the schema matching task we use the traditional well-known Purchase Order dataset~\cite{DO2002a}. The schemata used from this dataset %are medium size, with 142 and 46 attributes, and with 
contain high information content (data types and instance examples) to be presented (see Figure~\ref{fig:inteface} for example).

\noindent\textbf{Entity Matching:}\footnote{\url{https://agp.iem.technion.ac.il/incognitomatch/ER/}} For the entity matching task we use a Beer structured dataset~\cite{beer} from Magellan~\cite{konda2016magellan}. The entities in this dataset are composed of four elements (please refer to Figure~\ref{fig:example} and~\ref{fig:inteface}) which are available to the user.

\noindent\textbf{Text Matching:}\footnote{\url{https://agp.iem.technion.ac.il/incognitomatch/QQP/}} For the text matching task we use the quora question pairs dataset~\cite{qqp}, which is included in the GLUE benchmark~\cite{wang2018glue}. For this domain\footnote{We use domain and task interchangeably, refering to diffrent matching tasks.} the user is provided solely with the question text.

\subsection{Human Matchers}\label{sec:participants}

We use Prolific Academic\footnote{\url{https://www.prolific.co/}} as a crowdsourcing application from which the participants are directed to IncognitoMatch. We accepted only users using a desktop (no mobile devices) with English as a first language and at least 95\% approval rate. Each participant was provided with 30-50 matching decisions, each annotated with decision time and confidence. For privacy and confidentiality, participants' personal information (age, gender, etc.) is not used or shared. For each dataset we gathered 100 valid participants by screening low performance participants and experiments that had technical issues and errors (e.g., the user left the task in the middle). The full dataset consists of about 12,000 matching decisions.
 
On average a participant spent $\sim$15 minutes on the experiments out of which $\sim$5 minutes spent on instructions and riddles and $\sim$10 minutes on answering matching questions. A typical decision time is between 5 and 15 seconds (average of 11.5 seconds) with an average reported confidence of 0.59 and smoothed confidence (absolute confidence level disregarding the binary decision, see Eq.~\ref{eq:sconf}) of 0.5. This observation is different than the setting of open-ended crowdsourcing~\cite{ackerman2019cognitive,shraga2020learning,shraga2021powarematch} where most participants had more extreme confidence levels (close to 1 for match and close to 0 for no-match). Average decision times regarding a SM sample task are the longest among the tasks (14.5 seconds) and are taken with the lowest average confidence (0.48). An average ER decision is taken faster (10 seconds)  than an average TM decision (12 seconds), but with lower confidence (0.6 for EM and 0.65 for TM).      

\begin{figure}[h]
	\centering
	\includegraphics[width=.45\textwidth]{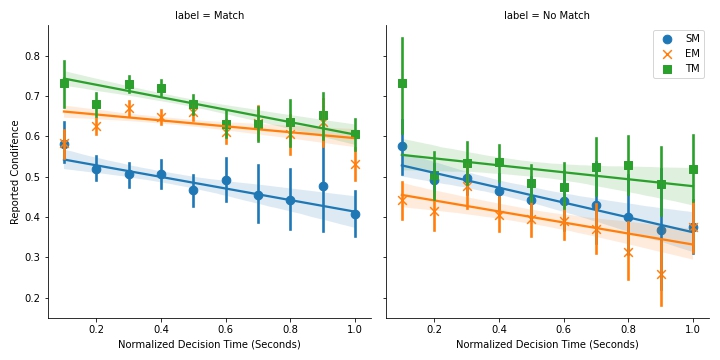}
	\caption{Time vs. Confidence for each Matching Task}
	\label{fig:time_conf}
\end{figure}

Figure~\ref{fig:time_conf} provides an analysis of the relationship between (normalized) decision times and (bucketed) reported confidence, similar to the one presented in~\cite{ackerman2019cognitive}, by matching task. It is quite clear that a negative correlation between the two is also present in our setting, especially for the task of schema matching. In general (using all the decisions from all tasks) a negative statistically significant correlation of $-.103$ (p-val$<0.001$) is observed between decision times and reported confidence. As illustrated in Figure~\ref{fig:time_conf}, the highest (absolute) correlation exist in SM ($-.098$). Milder correlation is observed for EM ($-.091$) and TM ($-.084$). All reported correlations are statistically significant with p-val$<0.001$. One factor that may explain the difference between the different tasks is the context provided to the user. The decision time encodes the exploration of the provided context which is far richer in the SM, reasonably present for EM and does not exist for TM. We note that our feature analysis (to be presented in Section~\ref{sec:ablation}) suggests that, different from the observation by Ackerman et al.~\cite{ackerman2019cognitive}, decision times are less predictive when it comes to correctness.

\subsection{Experimental Methodology}\label{sec:method}

The overall methodology assumes the existence of some real ground truth labels, i.e., annotated decisions of users. This assumption is realistic since existing benchmarks usually include labels that are validated and corrected over the years. Using such a set, we train a model that can later be used to calibrate the labels provided for unseen decision (what is referred to as a test set). Since we experiment with different types of tasks (domain), we create 4 cross validation settings, varying over three dimensions of train-test splits (Table~\ref{tab:TrainTest}), namely same domain, same samples, and same users.

\noindent\textbf{V1:} A na\"ive approach would be to split the users that solved the same matching task (same domain) with the same set of samples into a training set and test set. In other words this simplifying variation assumes we aim to \emph{calibrate new users on a known task}. In this case the majority voting feature (see Section~\ref{sec:ML}) must be excluded since the same samples are used. 

\noindent\textbf{V2:} We fix the users in the domain and change the the sample set, i.e., we now have \emph{known users solving a different task from the same domain}. 

\noindent\textbf{V3:} Here we still use the same matching task (domain) for both training and testing; yet we now also see \emph{new users solving a diffrent task from the same domain}. 

\noindent\textbf{V4:} The most challenging setting is V4 where in testing time we see \emph{new users solving a diffrent task from a diffrent domain}

In all settings we use 70\% of samples for training and 30\% were reserved for testing (over which we report the results).

\begin{table}[H]
	\caption{Cross Validation Settings}
	\label{tab:TrainTest}
	\centering 
	\begin{tabular}{l c c c}
		\multicolumn{1}{r}{\textbf{Setting}$\rightarrow$} & \multirow{2}{*}{Same Domain}  & \multirow{2}{*}{Same Samples} & \multirow{2}{*}{Same Users}\\
		$\downarrow$\textbf{Name} & & & \\
		\toprule
		V1 & \cmark & \cmark & \xmark\\
		V2 & \cmark & \xmark & \cmark\\
		V3 & \cmark & \xmark & \xmark\\
		V4 & \xmark & \xmark & \xmark\\
	\end{tabular}
	%				\vspace{-.25cm}
\end{table}

%To better explain the variations, assume, for simplicity, that we have two schema matching tasks SM1 and SM2 and one entity matching task EM1. In addition we have four sets of users dubbed U0, U1, U3 and U4, such that U0 and U1 solved SM1, U3 solved SM2, and U4 solved EM1. If we use U0, in this simplified setting, as the training set, U1, U3, and U4 would be used as a test set for variations V1, V3, and V4, respectively. If we also assume that U0 also solved SM2, then V2 would mean using their decisions over SM1 for training and their decisions for SM2 as a test set.

\noindent\textbf{Implementation:} We use python's Sklearn\footnote{\url{https://scikit-learn.org/}} for the implementation of the classifiers. Specifically, we trained ten different classifiers\footnote{KNN, Linear SVM, RBF SVM, Decision Tree, Random Forest, FC Neural Net, AdaBoost, Naive Bayes, QDA, and SGD} out of which we chose the best performing one over the training set to apply over the test set. Reported results are averaged over 20 different runs for each model.

\noindent\textbf{Metrics:} Different from traditional matching evaluation, in our setting we are equally interested in true positives and true negatives. The reason for that is the objective of this study, that is, to account for the human in-the-loop. Specifically, in our setting, we assess the performance using the accuracy measure, i.e., the proportion of correctly labeled samples out of all samples. The intuition is that we want to evaluate the ability to correctly label both match (true positives) and non-match (true negatives) decisions.

%Specifically, we are not considering the task as a validation task where usually the true positives are more important. Rather, we want the human to be able to accurately label the matching sample. In what follows we assess the performance using accuracy measure, i.e., the proportion of correctly labeled samples out of all samples.

\noindent\textbf{Baseline:} For the current study we are interested in providing a proof-of-concept for calibrating human labels. In what follows, our baseline is defined as $l_i$ (see Section~\ref{sec:model}), i.e., the binary label provided by the user. In the IncognitoMatch setting, using $l_i$ means that the reported confidence is greater than 0.5, which is also equivalent to a likelihood-based baseline~\cite{shraga2021powarematch}. We also note that we pre-screen low performing matchers. An alternative would be to consider the screening as an additional baseline as in related works~\cite{callison2009fast,ipeirotis2010quality,daniel2018quality}.

\subsection{Results (Over V1-V4)}\label{sec:res}

We now analyze the effectiveness of our approach. Table~\ref{tab:res} provides an aggregated comparison between our method and the baseline by cross validation setting (see Table~\ref{tab:TrainTest}).

\begin{table}[H]
	\caption{Results per Setting (Aggregated)}
	\label{tab:res}
	\centering 
	\begin{tabular}{l|c|c}
		\textbf{Setting} & Baseline & Our Method (HumanAL)\\
		\toprule
		V1 & 0.751 & \textbf{0.766} (+2.0\%) \\
		V2 & 0.744 & \textbf{0.813} (+9.3\%)\\
		V3 & 0.750 & \textbf{0.903} (+20.4\%) \\
		V4 & 0.751 & \textbf{0.787} (+4.8\%) \\
	\end{tabular}
\end{table}

Our suggested method is able to provide improvements over all settings. Among settings, V1 and V4 demonstrated the lowest improvement rates over the baseline (2\% and 4.8\%, respectively). This observation is not surprising regarding V4 as the most challenging setting where we examine \emph{cross-domain} abilities. In fact, a 4.8\% cross-domain improvement is quite encouraging and shows that modeling human behavior can be generalized beyond a single (matching) domain. In the case of V1, the ablation study (see Section~\ref{sec:ablation}) will reveal the majority voting as a very important feature set, which is excluded in this setting (see Section~\ref{sec:setup}). V3 demonstrated the most improvement, which can be attributed to its ability to generalize (new users and samples yet same domain). In what follows, \emph{the results suggest that if we want to extend, for example, the QQP dataset, training our method on the existing QQP benchmark can provide high quality labels for new (unseen) data, even if the initial labels are not that accurate.}

\begin{figure}[t]
	\centering
	\includegraphics[width=.45\textwidth]{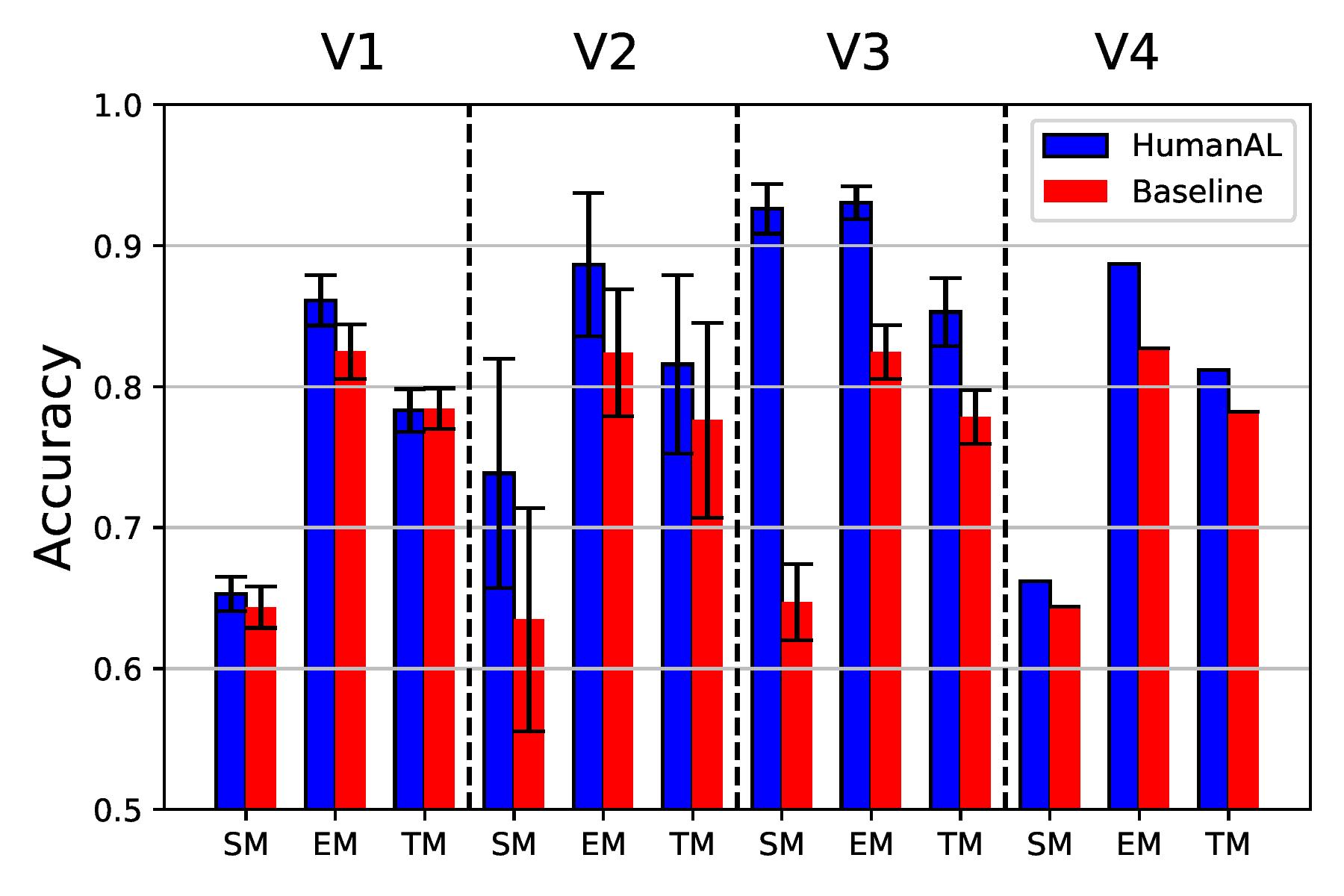}
	\caption{Results per Setting for Each Domain Seperatly.}
	\label{fig:accuracy_all}
\end{figure}

We now analyze the results by domain using Figure~\ref{fig:accuracy_all}. For V4, the indicated domain is the one used for testing. For example, the bars for SM represent that the model was trained on EM and TM and tested on SM. 

Unsurprisingly, on average, baseline human performance (red bars) over SM (0.642) is lower than over EM (0.825) and TM (0.781). The difference between EM and TM is quite surprising and can be explained by the similarity between tasks and fact that EM provides additional context. The largest average improvement (+38.2\%) is also achieved in the EM domain and smallest for the TM domain (+17.7\%). An interesting peak is observed for V3 setting in the SM domain where HumanAL provide 79.5\% improvement, while only 1\% improvement is achieved in the V1 setting for the TM domain. Finally, the largest variance is observed in the V2 setting which may be associated with varying complicatedness of samples within the same domain.

We now turn our efforts to analyze the feature space over the most challenging setting of V4.

\subsection{Ablation Study (Over V4)}\label{sec:ablation}

\begin{figure}[h]
	\begin{subfigure}{.45\linewidth}
		\centering
		\caption{``Isolating'' Features}
		\includegraphics[width=\linewidth]{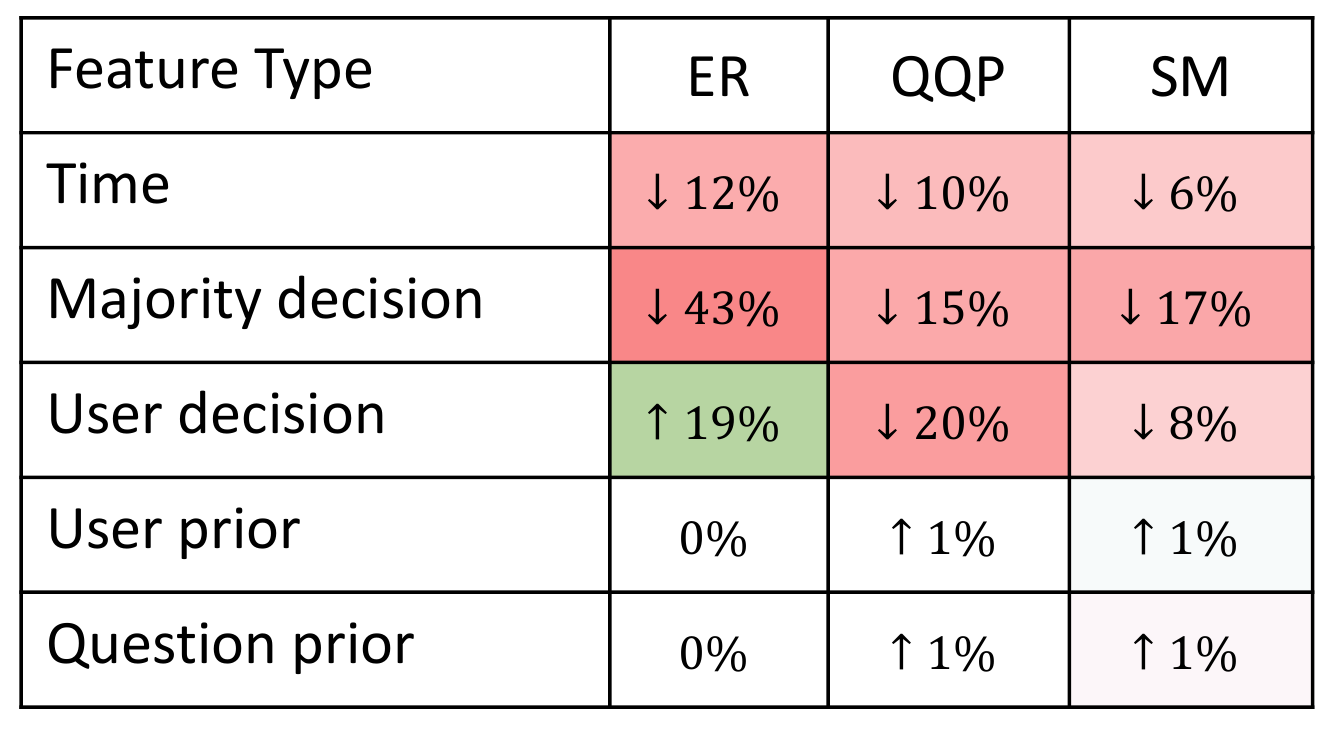}
		\label{fig:ablation_iso}
	\end{subfigure}
	\begin{subfigure}{.45\linewidth}
		\centering
		\caption{``Dropping'' Features}
		\includegraphics[width=\linewidth]{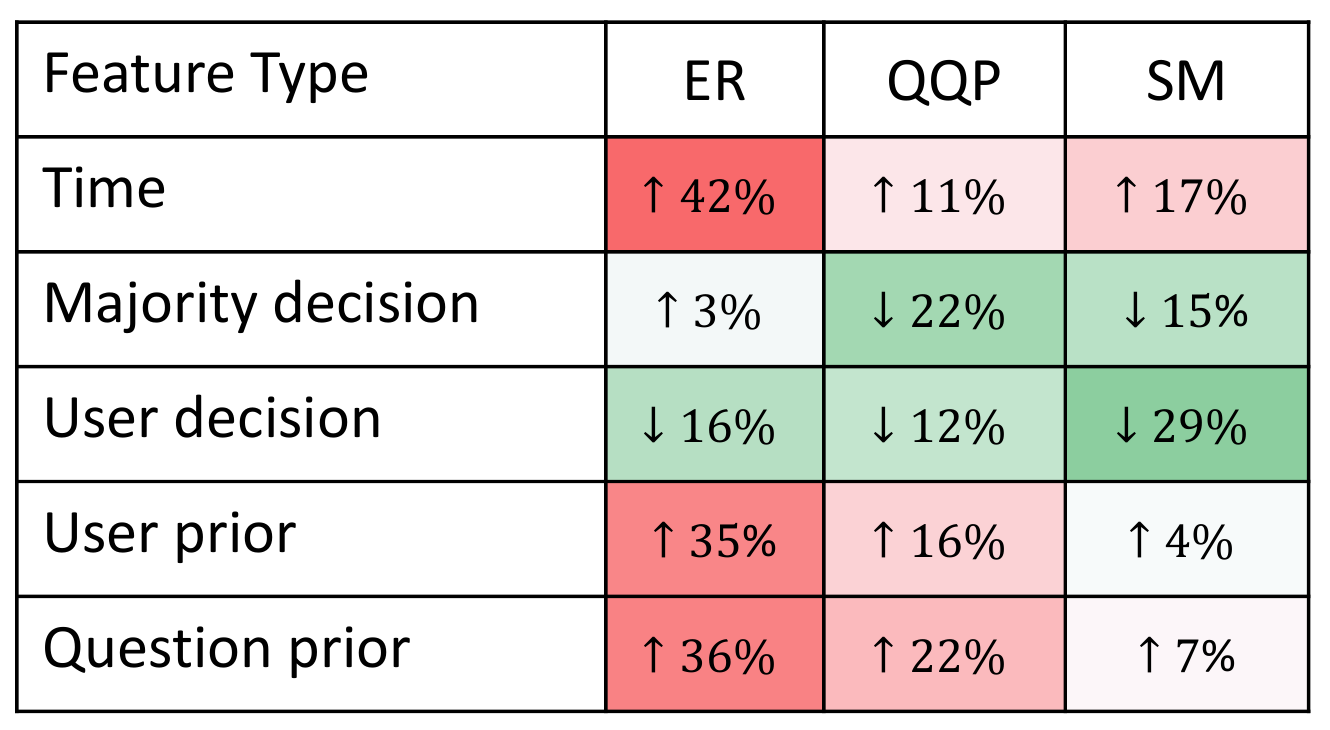}
		\label{fig:ablation_drop}
	\end{subfigure}
	\centering
	\caption{Ablation Study of the feature set using a isolating (a) and dropping (b) methodology. The indicated percentages are improvements over the baseline.}
	\label{fig:main}
\end{figure}

We analyze the importance of features as sets using an ``isolating'' and ``dropping'' methodology. For the former (Figure~\ref{fig:ablation_iso}) we isolate each feature set and only use its features for training. For the latter (Figure~\ref{fig:ablation_drop}) we exclude a feature set from training. \emph{Higher} accuracy in isolation means \emph{more important} feature set while for dropping, \emph{lower} accuracy means \emph{more important} feature set. We use a color coding (green for improvement and red for decrease) to enhance readability.

When dropping features (Figure~\ref{fig:ablation_drop}),  Majority decision and User decision are significant as the results decrease when excluding them. On the other hand, time is not significant and may actually decrease the accuracy. This observation comes as a surprise since previous works indicated its importance~\cite{ackerman2019cognitive,shraga2020learning,shraga2021powarematch}. The main difference between ours that previous works setting is the crowdsourcing setup (here we focus on boolean instead of open-ended). In what follows, as the decisions are not \emph{elective} (found and selected by users), the time for a given decision is less significant. Also, we can see that the priors are not as important. Looking at isolating features, we observe that none of the sets is able to perform well independently. We note that although it seems that the priors work well, they simply learn to predict the binary label (same as the baseline), thus they do not provide any additional meaningful information. 
\section{Discussion and Future Directions}\label{sec:con}

In this work, we presented a novel approach to treat humans input, focusing on the labels they provide. We showed that using a cognitive-aware feature extraction of decisions, we can generate better labels. Our experiments further suggested that given a new (matching) dataset to be labeled, the most significant improvement is achieved when a model is pre-trained in the same domain (different users and samples). Moreover, the experiment also indicated HumanAL ability to support cross-domain labeling such that if one's pre-trained model happens to be trained on a different domain they would still benefit from increased label quality.

The main idea of this paper is to provide a vision regarding the way human input should be addressed. As an early stage research, there is still room for improvement and future works. One straight-forward idea is to also include non-matching domains, for example, classification tasks, for which, we believe, the observations from this paper can applied to. A recent study~\cite{northcutt2021pervasive} highlighted that even well-known benchmarks suffer from label errors, and we believe that using our methodology for the next generation of benchmark labeling might reduce such errors.  

We would also like to extend and polish the feature sets to better reflect human decision making process. For example, IncognitoMatch, also includes mouse movements which were shown to be predictive to determine expertise~\cite{shraga2020learning}. Another interesting line of research would be setting more demanding prerequisites for the participants. Some previous works, e.g., Magellan~\cite{konda2016magellan,doan2017human}, use undergrad and grad students as labelers, while in this paper (and more others) the focus is on crowd workers. We initiated an experiment with varying expertise levels (crowd workers, pre-screened crowd workers, undergrads, grads, researchers) to examine these differences. 

Finally, with IncognitoMatch we can also control the context presented to the user. Another future direction would be the examine this context and its assistance to users in completing the task.  

\section*{Acknowledgments}
This work was supported in part by the National Science Foundation (NSF) under award numbers IIS- 1956096. We thank Gal Goldstein for his assistance in conducting the empirical evaluation and Coral Scharf for IncognitoMatch maintenance.

%- More than one matching system? 
%- Participants background (here refer to the initial study)
%- beyond matching
%- 

%%% The next two lines define the bibliography style to be used, and
%%% the bibliography file.
\bibliographystyle{ACM-Reference-Format.bst}
\balance
\bibliography{ltsLong}

\end{document}